\documentclass[aps,prd,amsmath,eqsecnum,showpacs]{revtex4}
\usepackage{bm}
\usepackage{graphicx}
\begin{document}
\title{Numerical investigation of friction in inflaton equations of motion}
\author{Ian D. Lawrie}
\email{i.d.lawrie@leeds.ac.uk}
\affiliation{School of Physics and Astronomy, The University of Leeds, Leeds LS2 9JT, England.}
\date{\today}
\begin{abstract}
The equation of motion for the expectation value of a scalar quantum field does not have
the local form that is commonly assumed in studies of inflationary cosmology.  We have recently argued
that the true, temporally non-local equation of motion does not possess a time-derivative expansion and
that the conversion of inflaton energy into particles is not, in principle, described by the
friction term estimated from linear response theory.  Here, we use numerical methods to
investigate whether this obstacle to deriving a local equation of motion is purely formal, or of
some quantitative importance.  Using a simple scalar-field model, we find that, although the
non-equilibrium evolution can exhibit significant damping, this damping is not well described
by the local equation of motion obtained from linear response theory.  It is possible that linear
response theory does not apply to the situation we study only because thermalization turns out to be
slow, but we argue that that the large discrepancies we observe indicate a failure of the local
approximation at a more fundamental level.
\end{abstract}
\pacs{11.10.Wx, 05.30.-d,98.80.Cq}
\maketitle
\section{Introduction\label{intro}}

The dynamics of a classical inflaton field is commonly assumed to be governed by a local equation of
motion of the form
\begin{equation}
\ddot{\phi} + 3H\dot{\phi} + \eta(\phi)\dot{\phi}+V_{\mathrm{eff}}(\phi)=0,
\label{localeom}
\end{equation}
where the friction term $\eta(\phi)\dot\phi$ represents the dissipative mechanism through which
inflaton energy is converted into particles.  In particular, in `warm inflation' scenarios
\cite{moss1985,berera1995,bererafang1995,berera2000}, this term is taken to be large
enough for a significant fraction of the energy to be dissipated during inflation.  Assuming that the
classical inflaton $\phi$ is the expectation value of quantum scalar field $\Phi$, its equation
of motion can be derived straightforwardly from a suitable quantum field theory, and it is \textit{not}
of the form (\ref{localeom}).  For example, if $\phi$ couples to a scalar quantum field $\chi$
through an interaction ${\textstyle\frac{1}{2}}g\phi^2\chi^2$, one obtains a term
$g\phi\langle\chi^2\rangle$ in the equation of motion, and the expectation value depends on $\phi$
through the effective mass of the $\chi$ particles, $m^2_\chi(\phi)=m_\chi^2+g\phi^2$.  The field
$\chi$ here might be the quantum part of $\Phi$ (that is, $\Phi=\phi+\chi$, with $\phi=\langle\Phi\rangle$)
or it might be an independent field.  In any case, the expectation value $\langle\chi^2\rangle$ is
a highly \textit{nonlocal} quantity: it depends in a complicated way on the entire history of $\phi$, in
contrast to the local equation (\ref{localeom}), where $\phi$, $\dot{\phi}$ and $\ddot{\phi}$ are all
evaluated at a single time $t$. This situation is quite generic:  a Yukawa coupling $g'\phi\bar\psi\psi$
to a spin-$\frac{1}{2}$ field $\psi$ produces a $\phi$-dependent mass for the $\psi$ particles and a term
$g'\langle\bar\psi\psi\rangle$ in the equation of motion, and in general the equation of motion
contains a sum of terms of this sort, but no friction term.

The question naturally arises, whether the true equation of motion might be adequately approximated,
under suitable circumstances, by a local equation similar to (\ref{localeom}).  An apparently rather
plausible argument suggests that this is indeed possible when $\phi$ changes sufficiently slowly
with time, and supplies an estimate of the friction coefficient $\eta(\phi)$.  To evaluate
$\langle\chi^2(t)\rangle$ at a \textit{fixed} time $t$, one expresses $\phi(t')$ at some earlier time $t'$ as
\begin{equation}
\phi(t')=\phi(t)+(t'-t)\dot\phi(t)+
\frac{1}{2}{\hbox{$(t'-t)^2$}}\ddot\phi(t)+\ldots.\label{phiexpansion}
\end{equation}
The evolution of $\chi(t')$ is governed by a Hamiltonian $H(t')=H(\chi,\phi(t'))$ that depends explicitly
on time through $\phi(t')$. Inserting the expansion (\ref{phiexpansion}), we have
\begin{equation}
H(t')=H(\chi,\phi(t)) +\Delta H(t'),
\end{equation}
where the first term, say $H_0(\chi)\equiv H(\chi,\phi(t))$ is independent of $t'$. Assuming that
the state of the $\chi$ field is always close to thermal equilibrium, the time-dependent correction
$\Delta H(t')$ can be treated using the standard methods of linear response theory, which yields the
approximation
\begin{equation}
\langle\chi^2(t)\rangle\approx\langle\chi^2\rangle^{\mathrm{eq}}+\mathrm{i}\int^t\mathrm{d}t'
\langle[\Delta H(t'),\chi^2(t)]\rangle^{\mathrm{eq}}
\end{equation}
where the expectation values are taken in the equilibrium ensemble associated with $H_0$.  Keeping just the
first two terms in (\ref{phiexpansion}), we find that $\Delta H(t')$ is proportional to $\dot\phi(t)$,
and in this way we recover an equation of the form (\ref{localeom}).  Using this or equivalent methods,
estimates of the friction coefficient were found in the 1980s for a single self-interacting scalar field
\cite{hosoya84,morikawa84,morikawa85}, and more recently (mainly in the context of warm inflation) for
more complicated models \cite{gleiser94,berera98,berera01,berera03,berera04,ramos04}.

Plausible though these calculations seem to be, they rely on an unproven assumption: namely, that the true
non-equilibrium dynamics can legitimately be replaced by the linear response of a state of exact thermal
equilibrium.  It is therefore important to ask whether the same results can be recovered from the
slow-evolution limit of the non-equilibrium dynamics.  Recently \cite{lawrie2002,lawrie03}, we have addressed
this question by exhibiting a sequence of approximations through which the exact non-equilibrium dynamics can
be reduced to a set of \textit{local}, coupled evolution equations for $\phi(t)$ and quantities characterizing
the state of the particles into which $\phi$ decays.  To recover the local equation of motion (\ref{localeom}),
it is necessary that the solution of these evolution equations should have an adiabatic expansion in powers
of derivatives of $\phi(t)$.  Somewhat surprisingly, we found, as reviewed in section \ref{eveqns} below,
that no such solution exists.  Formally, this means that the equation of motion for $\phi(t)$ is intrinsically
non-local, and does not have a time-derivative expansion.  In particular, the friction coefficient $\eta(\phi)$
does not exist.

This formal result indicates that the local equation of motion is never strictly valid, but that does not
necessarily mean that it cannot serve as a reasonably accurate approximation.  Intuitively, one might
still expect that (at least under favourable circumstances) dissipative effects, which certainly \textit{are}
a feature of the non-equilibrium evolution, should be fairly well represented by the friction term calculated
in linear response theory. It is this issue that we attempt to address here by numerical methods.  More
specifically, one can envisage a further approximation to the non-equilibrium evolution equations in which
self-energies are replaced with the values they would have in a state of thermal equilibrium.  Then
(see section \ref{eveqns} below) the solution of the resulting equations \textit{does} have a time-derivative
expansion, and one recovers essentially the friction coefficient obtained from linear response theory.
In what follows, we try to assess (within a particular model) whether the local equation of motion obtained
in this way provides a good approximation to the non-equilibrium dynamics.

Section \ref{eveqns} describes the non-equilibrium evolution equations and the adiabatic approximation
to these equations that we solve numerically.  In section \ref{results}, we present solutions for
the evolution from two different initial states:  one is a state in which the adiabatic approximation
predicts underdamped motion (section \ref{ud}), while in the other it predicts overdamped motion
(section \ref{od}).  In both cases, we find that, although although the non-equilibrium dynamics does
yield damping, this damping is not well described by the adiabatic approximation.  A brief account
of the numerical methods used to obtain these results is given in section \ref{numeric}.  Finally,
section \ref{discuss} is a self-contained summary, during the course of which consider what might
account for the failure of the adiabatic approximation in the situation we have studied.

\section{Approximate evolution equations for a scalar-field model\label{eveqns}}
As in \cite{lawrie2002,lawrie03} we work with a Minkowski-space theory.  Extending our analysis to a
Robertson-Walker spacetime presents no difficulties of principle, but it introduces complications which
seem to have no bearing on the issue we address.  Consider, then, a quantum field theory defined by the
Lagrangian density
\begin{equation}
\mathcal{L}={\textstyle\frac{1}{2}}\partial_\mu\Phi\partial^\mu\Phi-V(\Phi)+{\textstyle\frac{1}{2}}
\partial_\mu\chi\partial^\mu\chi-{\textstyle\frac{1}{2}}
m_\chi^2\chi^2-{\textstyle\frac{1}{4!}}\lambda\chi^4-{\textstyle\frac{1}{2}}g\Phi^2\chi^2.
\label{model}
\end{equation}
It should be emphasized at once that we will not attempt to study the exact non-equilibrium dynamics of
this model. In fact, we will arbitrarily discard various terms in its equations of motion which, though
they may well affect the dynamics significantly, are irrelevant to the question on which we wish to focus.
The first such approximation is to replace the quantum field $\Phi(\bm{x},t)$ with its expectation value
$\phi(t)$, where we assume a spatially homogeneous state.  Then the equation of motion for $\phi(t)$ is
\begin{equation}
\ddot\phi +V'(\phi)+g\phi\langle\chi^2(\bm{x},t)\rangle=0\label{phieom}
\end{equation}
while the dynamics of the remaining quantum field $\chi(\bm{x},t)$ is governed by the Lagrangian
\begin{equation}
\mathcal{L}_\chi={\textstyle\frac{1}{2}}
\partial_\mu\chi\partial^\mu\chi-{\textstyle\frac{1}{2}}
m^2(t)\chi^2-{\textstyle\frac{1}{4!}}\lambda\chi^4
\end{equation}
with $m^2(t)=m_\chi^2+g\phi^2(t)$.  An approximate treatment of this non-equilibrium dynamics yields,
for the expectation value in (\ref{phieom}) (which is independent of $\bm{x}$ for a homogeneous state)
\begin{equation}
\langle\chi^2(\bm{x},t)\rangle = \int\frac{d^3k}{(2\pi)^32\omega_k(t)}\,\left[1+2n_k(t)+2\mathrm{Re}\,\nu_k(t)\right],
\label{chisq}
\end{equation}
where $\omega_k=\sqrt{k^2+m^2(t)}$ and the functions $n_k(t)$ and $\nu_k(t)$, which characterize the state of the
system of $\chi$ particles, are solutions of the evolution equations
\begin{eqnarray}
\epsilon\partial_tn_k(t)&=&\alpha_k(t)-\Gamma_k(t)\left[1+2n_k(t)\right]
+\epsilon\frac{\dot{\omega}_k(t)}{\omega_k(t)}{\mathrm{Re}}\,\nu_k(t)\label{dndt}\\
\epsilon\partial_t\nu_k(t)&=&-2i\left[\omega_k(t)-i\Gamma_k(t)\right]\nu_k(t)-\alpha_k(t)
+\epsilon\frac{\dot{\omega}_k(t)}{2\omega_k(t)}\left[1+2n_k(t)\right].\label{dnudt}
\end{eqnarray}
In these equations, $\alpha_k(t)$ and $\Gamma_k(t)$ are given by
\begin{eqnarray}
\alpha_k(t)&=&\frac{c}{4\pi^2}\int d^3k_1d^3k_2d^3k_3
\frac{\delta(\omega_{k_1}+\omega_{k_2}-\omega_{k_3}-\omega_{k})
\delta(\bm{k}_1+\bm{k}_2-\bm{k}_3-\bm{k})}{\omega_k\omega_{k_1}\omega_{k_2}\omega_{k_3}}
\nonumber\\
&&\times\left[(1+n_{k_1})(1+n_{k_2})n_{k_3}+n_{k_1}n_{k_2}(1+n_{k_3})\right]
\label{alphadef}\\
\Gamma_k(t)&=&\frac{c}{4\pi^2}\int d^3k_1d^3k_2d^3k_3
\frac{\delta(\omega_{k_1}+\omega_{k_2}-\omega_{k_3}-\omega_{k})
\delta(\bm{k}_1+\bm{k}_2-\bm{k}_3-\bm{k})}{\omega_k\omega_{k_1}\omega_{k_2}\omega_{k_3}}
\nonumber\\
&&\times\left[(1+n_{k_1})(1+n_{k_2})n_{k_3}-n_{k_1}n_{k_2}(1+n_{k_3})\right]
\label{gammadef}
\end{eqnarray}
with $c=\lambda^2/64(2\pi)^3$, while $\epsilon$, which has the value $\epsilon=1$, is a formal
parameter introduced to facilitate an adiabatic expansion.

The approximations that lead to these equations are described in detail in \cite{lawrie03}.  Essentially,
they arise as a solution to the Dyson-Schwinger equations for 2-point functions, via a local \textit{Ansatz}
for self-energies, which are evaluated at 2-loop order in perturbation theory.  For a free-field theory,
with $\lambda=0$, which also means $\alpha_k(t)=\Gamma_k(t)=0$, the evolution equations are exact (see, for
example \cite{morikawa84}) and the functions $n_k(t)$ and $\nu_k(t)$ can be identified in terms of
creation and annihilation operators for $\chi$ particles:
\begin{eqnarray}
\langle a_k^\dag(t)a_{k'}(t)\rangle&=&(2\pi)^3\delta(\bm{k}-\bm{k}')n_k(t)\label{nkcan}\\
\langle a_k(t)a_{k'}(t)\rangle&=&(2\pi)^3\delta(\bm{k}+\bm{k}')\nu_k(t).\label{nukcan}
\end{eqnarray}
With interactions included, (\ref{dndt}) can be interpreted as a Boltzmann equation for number densities
$n_k(t)$;  the quantity
\begin{equation}
S_k(t)=\alpha_k(t)-\Gamma_k(t)[1+2n_k(t)]\label{sdef}
\end{equation}
is the standard scattering integral associated with two-body elastic scattering, while the source
\begin{equation}
j(t)=\frac{\dot{\omega}_k(t)}{\omega_k(t)}{\mathrm{Re}}\,\nu_k(t)\label{jdef}
\end{equation}
represents particle creation and absorption due to the time-dependent mass $m(t)$.

Although the evolution equations (\ref{dndt}) and (\ref{dnudt}) are local, their solution is not.
One can attempt to obtain a solution in the form of a time-derivative expansion, by writing
\begin{eqnarray}
n_k(t)&=&n_k^{\mathrm{eq}}(t)+\epsilon\delta n_k(t)+\mathrm{O}(\epsilon^2)\\
\nu_k(t)&=&\nu_k^{\mathrm{eq}}(t)+\epsilon\delta \nu_k(t)+\mathrm{O}(\epsilon^2),
\end{eqnarray}
expanding in powers of $\epsilon$ and finally setting $\epsilon=1$. However, it turns out that no
such solution exists, for the following reason.  To find the leading correction $\delta n_k$, one has to
solve the integral equation
\begin{equation}
\int_0^\infty dk'K(k,k')\delta n_{k'}=\partial_tn^{\mathrm{eq}}_k-\frac{\dot{\omega}_k}{\omega_k}
{\mathrm{Re}}\,\nu_k^{\mathrm{eq}},
\label{deltaneqn}
\end{equation}
where
$
K(k,k') = \left(\delta S_k/\delta n_{k'}\right)_{n=n^{\mathrm{eq}}}$.  It is straightforward to
show that conservation of energy in two-body scattering implies the sum rule
$ \int_0^\infty dk\,k^2\omega_kK(k,k')=0$.  The right-hand side of (\ref{deltaneqn}) does not respect
this sum rule, so the equation is not self-consistent and the desired solution does not exist.  (For
elastic scattering, there is a second sum rule associated with particle-number conservation, which is
also not respected by (\ref{deltaneqn}); however, corrections to the Boltzmann equation (\ref{dndt}) from
higher orders of perturbation theory would introduce inelastic scattering terms, destroying this second
sum rule.)

To the extent that our approximations afford an adequate description of non-equilibrium evolution (an
important caveat, upon which we comment further in section \ref{discuss}) this shows that the equation
of motion (\ref{phieom}) does not have a time-derivative expansion and cannot be represented by a local
equation such as (\ref{localeom}).  However, we can recover a time-derivative expansion by making a
further approximation.  The functions $\alpha_k(t)$ and $\Gamma_k(t)$ are related to quasiparticle
self-energies $\Sigma^{\pm}_k(t)$ (whose exact meanings are given in \cite{lawrie03}) by
\begin{eqnarray}
\alpha_k(t)&=&-\frac{i}{4\omega_k(t)}\left[\Sigma^{+}_{k}(t)
+\Sigma^{-}_{k}(t)\right]\label{alphapresc}\\
\Gamma_k(t)&=&-\frac{i}{4\omega_k(t)}\left[\Sigma^{+}_{k}(t)
-\Sigma^{-}_{k}(t)\right].\label{gammapresc}
\end{eqnarray}
Let us replace these self-energies with constant equilibrium values.  In what follows, we will call
this the \textit{adiabatic approximation}. Using this approximation, the time-derivative
expansion can be carried through without difficulty.  In particular, $\alpha_k(t)$ and $\Gamma_k(t)$
no longer depend on $\delta n_k(t)$, and the troublesome equation (\ref{deltaneqn}) does not arise.
Since $\dot\omega_k(t)$ is proportional to $\dot\phi(t)$, so are $\delta n_k(t)$ and $\delta\nu_k(t)$,
and these contributions to $\langle\chi^2\rangle$ give rise to a friction coefficient
\begin{equation}
\eta(\phi)=\frac{g^2\phi^2}{4\pi^2}\int_0^\infty dk\,k^2
\left[\frac{\beta(\omega_k^2+3\Gamma_k^2)n_k(1+n_k)}
{\Gamma_k(\omega_k^2+\Gamma_k^2)^2}+\frac{\Gamma_k(\omega_k^2-3\Gamma_k^2)(1+2n_k)}{\omega_k(\omega_k^2
+\Gamma_k^2)^3}\right].\label{etaeq}
\end{equation}
Here, $n_k$ stands for the equilibrium Bose-Einstein distribution $n_k^{\mathrm{eq}}
=\left(e^{\beta\omega_k}-1\right)^{-1}$, while $\Gamma_k$ is found by substituting this equilibrium
distribution in (\ref{gammadef}). This expression agrees, for example, with the result given in
\cite{morikawa85} in the limit $\Gamma_k<<\omega_k$ considered by these authors.  Roughly speaking,
the second term represents a loss of inflaton energy arising from particle creation, while the first
represents a transfer of energy to particles already present.

This agreement is not surprising:  the adiabatic approximation we have just introduced is roughly
equivalent to the assumptions on which linear response theory is based.  Intuitively, it would seem
that this should be a fair approximation if $\phi(t)$ does not change too rapidly, and if the state
of the $\chi$ particles does not stray too far from an equilibrium distribution.  We now attempt
to test this intuition by comparing numerically the non-equilibrium evolution resulting from
(\ref{dndt}) and (\ref{dnudt}) with that implied by the adiabatic approximation.

To simplify matters a little, we observe that the frictional effect we want to study arises from the
last two terms in the expression (\ref{chisq}) for $\langle\chi^2\rangle$, and will discard the first
term.  Thus, we study the equation of motion
\begin{equation}
\ddot\phi +V'(\phi)+g\phi\langle\chi^2(\bm{x},t)\rangle_\mathrm{trunc}=0\label{phieom2}
\end{equation}
with the truncated expectation value
\begin{equation}
\langle\chi^2(\bm{x},t)\rangle_\mathrm{trunc}
 = 2\int\frac{d^3k}{(2\pi)^32\omega_k(t)}\,\left[n_k(t)+\mathrm{Re}\,\nu_k(t)\right],
\label{chisqtr}
\end{equation}
the evolution of $n_k(t)$ and $\nu_k(t)$ being governed by (\ref{dndt}) and (\ref{dnudt}).  The
adiabatic approximation yields the estimate
\begin{equation}
g\phi\langle\chi^2(\bm{x},t)\rangle^{\mathrm{adiab}}_\mathrm{trunc}=\Delta V'(\phi)+\eta(\phi)\dot\phi
\label{chisqadiab}
\end{equation}
where $\eta(\phi)$ is given by (\ref{etaeq}) and the contribution from the leading order terms of the
adiabatic expansion is
\begin{equation}
\Delta V'(\phi)=\frac{g\phi}{2\pi^2}\int_0^\infty dk\,\frac{k^2}{\omega_k}
\left[n_k^{\mathrm{eq}}+\mathrm{Re}\,\nu_k^{\mathrm{eq}}\right]\label{deltavpr}
\end{equation}
with
\begin{equation}
\mathrm{Re}\,\nu_k^{\mathrm{eq}}=-\frac{\Gamma_k^2(1+2n_k^\mathrm{eq})}{2(\omega_k^2+\Gamma_k^2)}.
\label{nueq}
\end{equation}
In order to evaluate $\langle\chi^2(\bm{x},t)\rangle^{\mathrm{adiab}}_\mathrm{trunc}$, we need values
for the inverse temperature $\beta$ of the assumed equilibrium state.  To this end, we observe that
the evolution equations (\ref{phieom2}), (\ref{dndt}) and (\ref{dnudt}) conserve exactly the energy
given by
\begin{equation}
E=\frac{1}{2}\dot\phi^2+V(\phi)+\frac{1}{2\pi^2}\int_0^\infty dk\,k^2\omega_kn_k.\label{energy}
\end{equation}
We will determine the evolving temperature in the adiabatic approximation by requiring this energy to
remain constant.

To avoid confusion, it may be worthwhile to reiterate that our goal is to test whether the expectation
value (\ref{chisqtr}) is adequately represented by the adiabatic expression (\ref{chisqadiab}).  We will do
this by comparing solutions to the non-equilibrium evolution equations (\ref{phieom2}), (\ref{chisqtr}),
(\ref{dndt}) and (\ref{dnudt}) with the adiabatic approximation \textit{to these equations}, which
consists of the equation of motion
\begin{equation}
\ddot\phi+\eta(\phi)\dot\phi+V'(\phi)+\Delta V'(\phi)=0,\label{eomadiab}
\end{equation}
with $\eta(\phi)$ and $\Delta V'(\phi)$ given by (\ref{etaeq}) and (\ref{deltavpr}) respectively, and
$\beta$ determined by energy conservation.  The fact that we have arbitrarily discarded several terms
does not affect this comparison, though of course it does mean that we are not in a good position to
study the true dynamics of the original model (\ref{model}).  Equally, our approximations are based in
part on perturbation theory but the neglected higher-order terms do not affect the comparison, and
we will not be concerned with choosing parameter values so as to make these terms small.
\section{Numerical results\label{results}}
We have obtained numerical solutions of the evolution equations presented in the previous section, taking
the simplest available potential, namely
\begin{equation}
V(\phi)=\frac{1}{2}M^2\phi^2.
\end{equation}
This section presents two sets of solutions.  In the first, the motion of $\phi$ is an underdamped
oscillation;  in the second, the motion is predicted \textit{by the adiabatic approximation} to be
overdamped.  In each case, we solve, on the one hand the non-equilibrium equation of motion (\ref{phieom2})
together with (\ref{dndt}) and (\ref{dnudt}), and on the other hand the adiabatic approximation to this
equation (\ref{eomadiab}), starting from identical initial conditions.  At the end of the section, we
comment briefly on the numerical methods used.
\subsection{Underdamped motion\label{ud}}
\begin{figure}
\begin{center}
\includegraphics[bb = 0 0 269 220,scale=1.4]{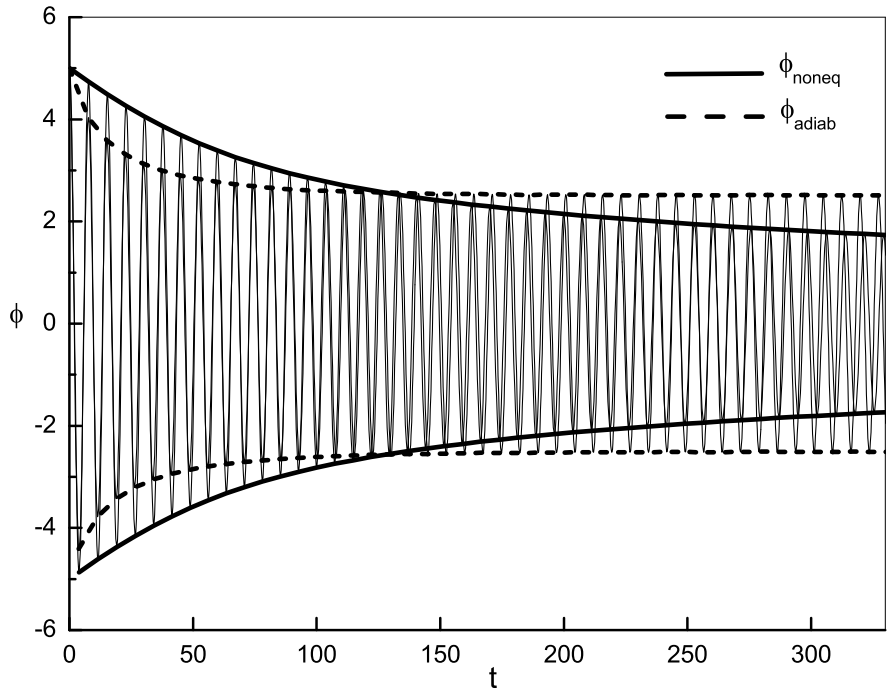}
\end{center}
\caption{\label{fig1}Underdamped motion of $\phi(t)$ as calculated from the non-equilibrium
evolution equations and via the adiabatic approximation.  Envelopes of the two oscillatory curves are
drawn to aid the eye.}
\end{figure}

Figure \ref{fig1} shows the motion of $\phi$ obtained in the two approximations, starting from a
more or less arbitrary initial state.  Throughout, the units are fixed by taking the mass of $\chi$ particles
to be $m_\chi=1$.  Here, we chose the parameter values $c=0.5$, $g=1.0$ and $M^2=0.4$.  The initial state
for $\phi$ was $\phi=5.0$, $\dot\phi=0$ and the initial state of the $\chi$ particles was a state of
thermal equilibrium with inverse temperature $\beta=0.3$.  With these values, the energy of the particles
is about 86\% of the total energy, so one might expect a significant transfer of further energy to these particles
as well as a frictional effect from particle creation.
The two superimposed oscillations in figure \ref{fig1} are not easily distinguished, so we have plotted
envelopes for the two curves: the solid envelope indicates the solution to the non-equilibrium evolution
equations, while the dashed one corresponds to the adiabatic approximation.  At first sight, there
seems to be a similar rate of damping in each case, but there are significant differences.  The friction
coefficient of the adiabatic approximation given in (\ref{etaeq}) is a rather complicated function of
$\phi$, but suppose that it can be written as $\eta(\phi) = \phi^2\hat\eta(\phi)$, where $\hat\eta(\phi)$
varies relatively slowly with $\phi$.  This suggests that large-amplitude oscillations should be strongly
damped, while at small amplitudes damping would be ineffective.  This behaviour is clearly apparent from
figure \ref{fig1}.  However, the solution of the full evolution equations shows a different tendency:
damping is smaller at large amplitudes than that predicted by the adiabatic approximation, and it does not
disappear at small amplitudes.

Thus, in this situation, although the non-equilibrium evolution equations do yield a damped motion, the
damping is not well represented by the friction term as estimated from the adiabatic approximation.
This is not necessarily unexpected, since both approximations yield fairly rapid oscillations, and the
adiabatic approximation might not be expected to work well.  We therefore turn to a different initial
state which, according to the adiabatic approximation, ought to result in overdamped motion and slow
evolution.
\subsection{Overdamped motion\label{od}}
Suppose, as above, that the adiabatic friction coefficient can be represented as $\eta(\phi)\approx\hat\eta\phi^2$
and also that the adiabatic correction to the effective potential (\ref{deltavpr}) is $\phi$ times
a slowly varying function.  Then the adiabatic equation of motion (\ref{eomadiab}) becomes
\begin{equation}
\ddot\phi+\hat\eta\phi^2\dot\phi+\mu^2\phi=0
\end{equation}
where $\hat\eta$ and $\mu^2=M^2+\Delta V'/\phi$ are, at least over some period of time, approximately constant.
If $\ddot\phi$ can be neglected, then this equation has the overdamped solution
\begin{equation}
\phi(t)\approx\left[\phi(0)^2-\frac{2\mu^2t}{\hat\eta}\right]^{1/2},\label{odsolution}
\end{equation}
with a ``terminal velocity''
\begin{equation}
\dot\phi\approx-\frac{\mu^2}{\hat\eta\phi}\,.\label{vterm}
\end{equation}
Under these circumstances, we have $\ddot\phi\approx-\mu^4/\hat\eta^2\phi^3$, so overdamped motion should
occur if $\vert\ddot\phi\vert\ll\vert\mu^2\phi\vert$, or
\begin{equation}
\left\vert\frac{\mu^2}{\eta^2}\right\vert\ll 1.\label{odcondition}
\end{equation}
Intuitively, the adiabatic approximation ought to be good if, in addition to this condition for neglecting
$\ddot\phi$, the state of the $\chi$ particles thermalizes rapidly.  According to the Boltzmann-like equation
(\ref{dndt}), the mode $n_k$ of the particle distribution relaxes towards equilibrium with a characteristic
relaxation time $\tau_k=1/\Gamma_k$.  Taking the $k=0$ mode as representative, one can guess at a rough
criterion for efficient thermalization, namely that the fractional change in the effective particle mass
$m^2=m_\chi^2+g\phi^2$ over a time $\tau_{k=0}$ should be small.  That is,
$\tau_{k=0}\vert dm^2/dt\vert\ll m^2$ or
\begin{equation}
\left\vert\frac{2g\phi\dot\phi}{m^2\Gamma_{k=0}}\right\vert\ll 1.\label{thermalization}
\end{equation}
Ideally, we would like to study a state in which both conditions (\ref{odcondition}) and (\ref{thermalization})
are well satisfied, but this proves to be quite difficult.  In particular, the second condition seems
hard to achieve.  Using (\ref{vterm}), it can be rewritten as
\begin{equation}
\left\vert\frac{2g\mu^2}{m^2\hat\eta\Gamma_{k=0}}\right\vert\ll 1
\end{equation}
and the thermal effects that make $\hat\eta$ and $\Gamma_k$ large also tend to produce large values of
$\mu^2$.  The dependence of $\eta(\phi)$ (equation (\ref{etaeq})) on the parameters at our disposal is
so complicated as to make a systematic survey of the parameter space impractical.  By trial and error, we have
chosen a set of values which yield
\begin{equation}
\left\vert\frac{\mu^2}{\eta^2}\right\vert\approx 5.9\times 10^{-5}\qquad\qquad\hbox{and}\qquad\qquad
\left\vert\frac{2g\phi\dot\phi}{m^2\Gamma_{k=0}}\right\vert\approx 0.97.\label{initcriteria}
\end{equation}
This means that \textit{according to the adiabatic approximation}, the system is strongly overdamped, but
thermalization is only moderately efficient.  A careful, but by no means exhaustive, exploration
suggests that this situation cannot be much improved within the model studied here.  The parameter values
leading to this state are $c=g=1.0$, $M^2=0.005$, $\beta = 0.09$ and $\phi = 80.0$.  The terminal velocity
(\ref{vterm}) turns out to be $\dot\phi=-0.08967\ldots$.

\begin{figure}
\begin{center}
\includegraphics[bb=0 0 273 221,scale=1.4]{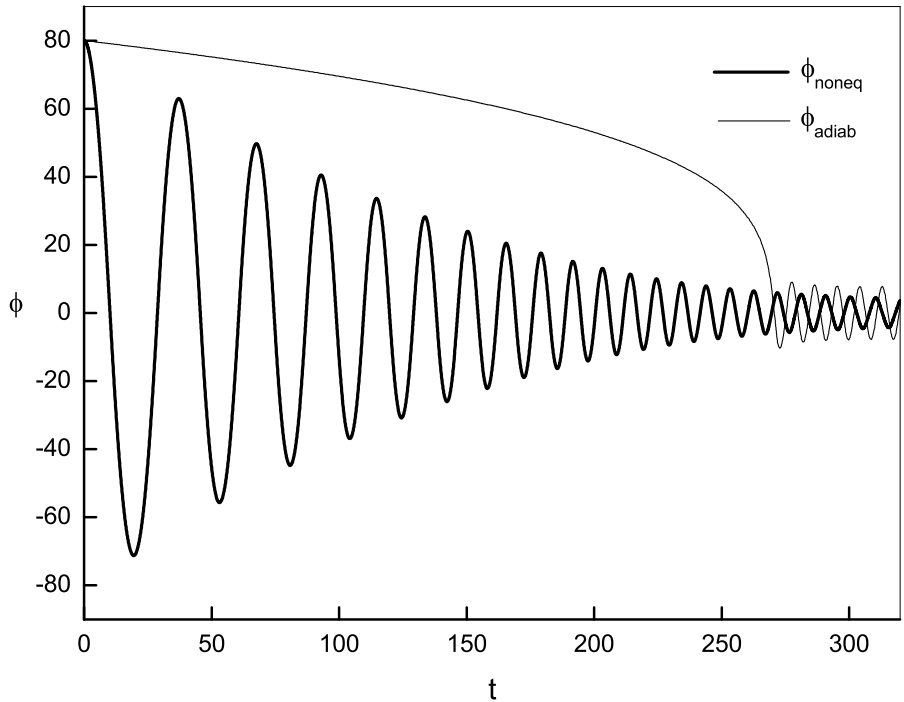}
\end{center}
\caption{\label{fig2}Motion of $\phi(t)$ as calculated from the non-equilibrium
evolution equations and via the adiabatic approximation.  As determined from the adiabatic approximation,
the initial conditions should lead to overdamped motion.}
\end{figure}

Taking these initial conditions, we find for $\phi(t)$ the motion depicted in figure \ref{fig2}.
Evidently, the motion obtained from the adiabatic approximation (the thinner curve) is indeed overdamped at
early times.  In fact the solution agrees well with (\ref{odsolution}) for times up to about $t=150$.  But
as $\phi$ becomes smaller, the condition (\ref{odcondition}) for overdamping no longer holds and, as in
figure \ref{fig1}, the motion at small amplitudes is essentially undamped.

\begin{figure}
\begin{center}
\includegraphics[bb=0 0 280 226,scale=1.4]{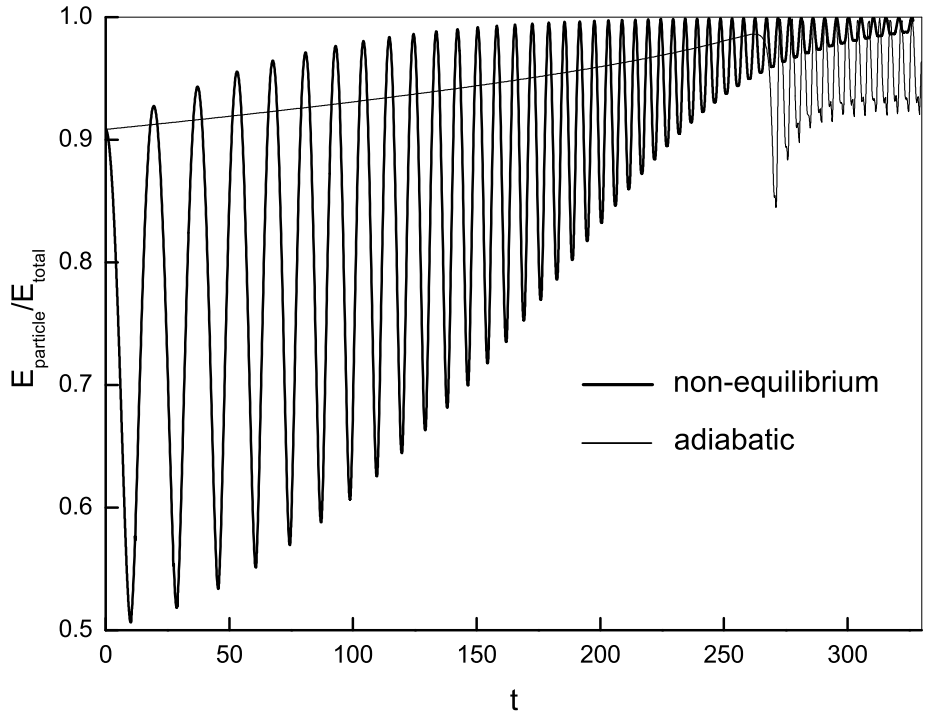}
\end{center}
\caption{\label{fig3}Energy of particles as a fraction of the conserved total energy.}
\end{figure}

The solution of the non-equilibrium evolution equations (the thicker curve in figure \ref{fig2})
is strikingly different:  although there is significant damping, the motion is far from being overdamped.
We observe, though, that the average rate of energy loss is quite similar, as indicated in figure \ref{fig3},
where we plot the energy that exists in the form of particles (the last term in (\ref{energy})) as a
fraction of the conserved total energy.  In fact, this figure shows that the overall conversion of inflaton
energy into particles is somewhat more efficient in the non-equilibrium evolution than is suggested by
the adiabatic approximation.

Of course, the question arises, what accounts for the large qualitative difference between the
non-equilibrium evolution and that obtained from the adiabatic approximation?  A simple answer
is that, as shown in section \ref{eveqns}, no adiabatic expansion of the non-equilibrium
evolution equations actually exists, and our numerical results merely confirm that the adiabatic
approximation is wrong.  This may perhaps be the appropriate answer, but it needs some further
discussion, because we also showed in section \ref{eveqns} that the adiabatic expansion
\textit{does} exist if we take the extra step of replacing self-energies with their equilibrium
values, and one might think that this step is relatively innocuous if the state is not too far
from equilibrium.

Consider first the initial state, which was chosen to be a state of exact thermal
equilibrium. According to the adiabatic approximation, the criterion for neglecting $\ddot\phi$
is very well satisfied (equation (\ref{initcriteria})), and the motion should be strongly overdamped.
However, we see from figure \ref{fig2} that $\vert\ddot\phi\vert$ as calculated from the non-equilibrium
evolution equations is very large.  The reason for this is not hard to discover.  In a state of
exact thermal equilibrium, the term $g\phi\langle\chi^2(\bm{x},t)\rangle_\mathrm{trunc}$ in the
equation of motion (\ref{phieom2}) is precisely the quantity $\Delta V'(\phi)$ that appears in
the adiabatic equation (\ref{eomadiab}). Thus, while the original equation of motion asserts that
$U(\phi)\equiv V'(\phi)+\Delta V'(\phi)$ is equal to $-\ddot\phi$, the adiabatic approximation
asserts that $U(\phi) \approx-\eta(\phi)\dot\phi$, the value of $\ddot\phi$ being very small.
Clearly, \textit{the adiabatic approximation gives a wrong answer for a state of exact thermal
equilibrium}. In fact, the informal arguments given in \cite{hosoya84,morikawa84} and reviewed
in \cite{lawrie03} show that the frictional effects arise precisely from small departures from
equilibrium.
\begin{figure}
\begin{center}
\includegraphics[bb=0 0 294 226,scale=1.4]{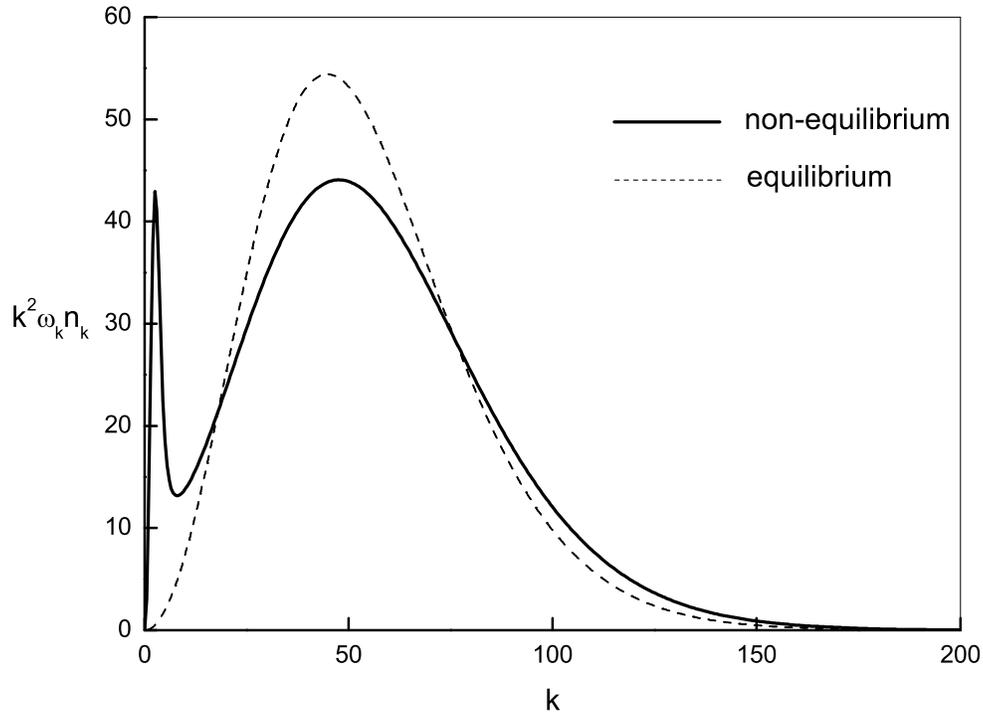}
\end{center}
\caption{\label{fig4}Weighted particle distribution $k^2\omega_kn_k$ at time $t=19.9$ (solid curve)
and the same quantity calculated from the equilibrium distribution for the same total particle
energy (broken curve).}
\end{figure}

\begin{figure}
\begin{center}
\includegraphics[bb=0 0 281 221,scale=1.4]{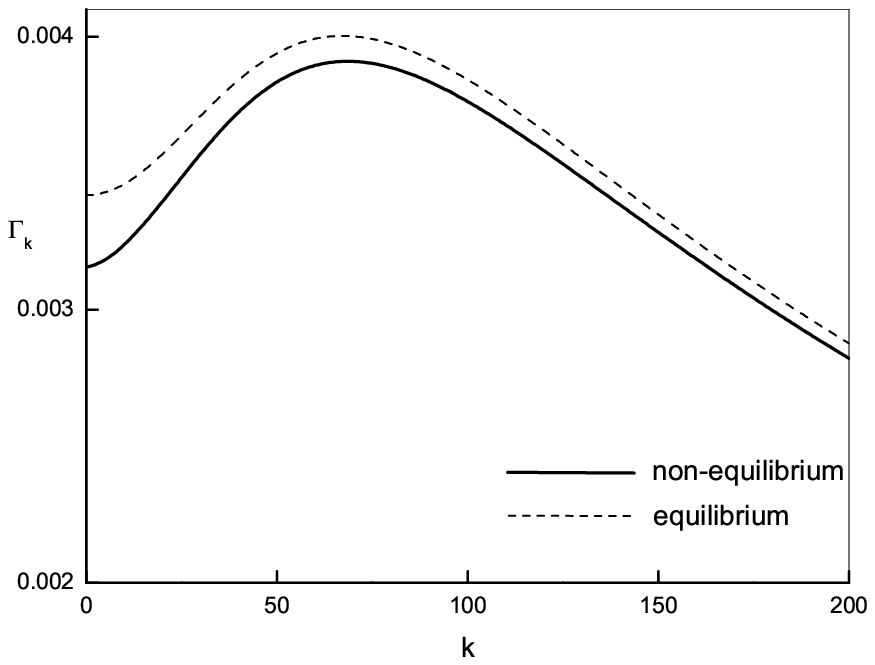}
\end{center}
\caption{\label{fig5}The quasiparticle width  $\Gamma_k$ calculated from the particle distributions of
figure \ref{fig4}.}
\end{figure}

Figure \ref{fig4} illustrates the departure from thermal equilibrium at times near $t\approx 20$,
corresponding to the first negative peak in figure \ref{fig2}, where about one half-cycle of
oscillation has been completed.  The quantity plotted is $k^2\omega_kn_k$, whose integral gives
the total energy in particles. The solid curve uses the distribution $n_k(t)$ generated by the
non-equilibrium evolution, while the broken curve uses the equilibrium distribution
$n^{\mathrm{eq}}_k$ with the temperature $\beta^{-1}$ adjusted to give the same total energy.
(This is quite independent of the equilibrium distribution $n_k^{\mathrm{adiab}}$ generated by
the adiabatic evolution.)  Clearly, these two distributions are different, but the difference
does not greatly affect the time evolution.  For example, we show in figure \ref{fig5} the
quasiparticle width $\Gamma_k$, which appears in the evolution equations (\ref{dndt}) and (\ref{dnudt}),
calculated from $n_k(t)$ and $n_k^\mathrm{eq}$, and the difference is seen to be marginal.  We can use
the effective value of $\beta$ calculated as above, together with the non-equilibrium values of $n_k$
and $\Gamma_k$ in the expressions (\ref{etaeq}) and (\ref{deltavpr}) to find an effective, non-equilibrium
value for the ratio $\mu^2/\eta^2$ which, according to (\ref{odcondition}), gives a criterion for
overdamping at times near $t=20$.  The result is that $\vert\mu^2/\eta^2\vert\approx 9.5\times 10^{-5}$.
Thus, the criterion for overdamping (as obtained from the adiabatic approximation) is well satisfied
by the non-equilibrium state at $t\approx 20$, just as it was by the initial equilibrium state.

The main point of the preceding discussion is this.  In order to obtain the adiabatic approximation,
we had to replace non-equilibrium self-energies $\Sigma_k[n]$ with the values $\Sigma_k[n^\mathrm{eq}]$
associated with a state of thermal equilibrium.  In the initial state at $t=0$ these two self energies
are identical, and in the state illustrated in figure \ref{fig5} the difference is quite small.
Clearly, this difference does not account for the drastic failure of the adiabatic approximation in the
situation we have examined.  The difference between the two evolutions in figure \ref{fig2} (and, for
that matter, those shown in figure \ref{fig1}) indicates that the expectation value
$\langle\chi^2(\bm{x},t)\rangle_\mathrm{trunc}$ is not well represented as the sum of the two terms
in (\ref{chisqadiab}), regardless of whether the coefficients $\Delta V'(\phi)$ and $\eta(\phi)$ are
calculated using the non-equilibrium state or an effective equilibrium state.  Thus, although the rate of
thermalization may well be an important factor, its importance does not lie in the relatively small
effect that inefficient thermalization would have on the values of these coefficients.

\begin{figure}
\begin{center}
\includegraphics[bb=0 0 287 226,scale=1.4]{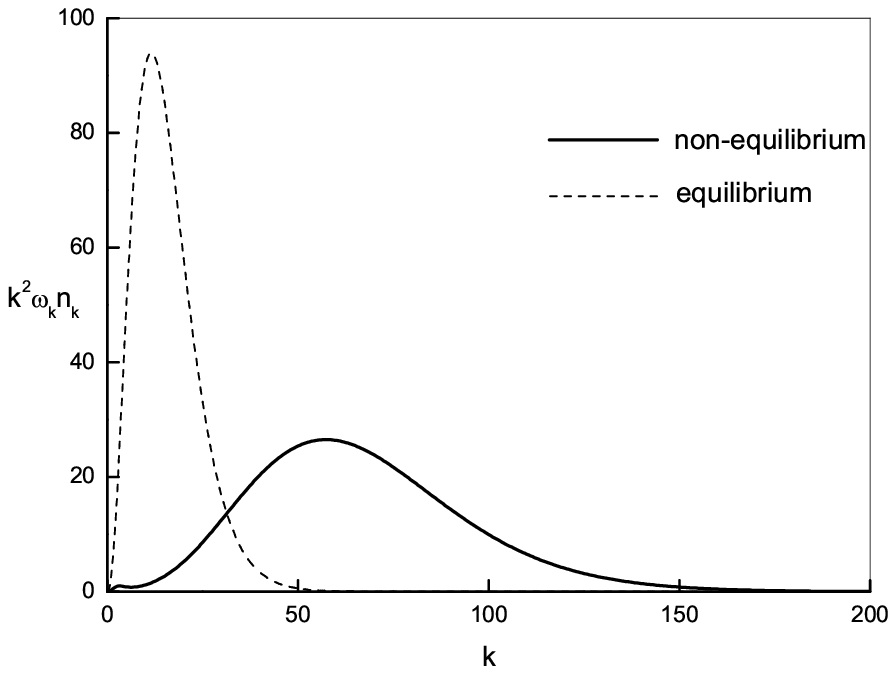}
\end{center}
\caption{\label{fig6}Weighted particle distribution at $t=10.2$, compared with the equilibrium distribution
for the same energy.}
\end{figure}

Having made this point, we defer further discussion to section \ref{discuss}, but we observe here
that thermalization in the situation we have studied is rather less efficient than one might
infer from the value $0.97$ for the ratio in (\ref{thermalization}).  In the initial state, the relaxation
time $\tau_k=\Gamma_k^{-1}$ has values $\tau_k\approx 400$ for all the modes used in the calculation,
and this is longer than the whole time for which the evolution was studied.  From the point of view
of the adiabatic approximation, this means that although thermalization might have been expected to be
moderately efficient by comparison with the initially slow evolution, it is much less efficient by
comparison with the more rapid evolution at later times.  In the non-equilibrium evolution, there are quite
large departures from equilibrium when $\phi$ changes rapidly.  For example, we show in figure \ref{fig6}
the particle distribution (again weighted by $k^2\omega_k$) at the time $t\approx 10.2$ where $\phi$ first
becomes very small.
This figure illustrates the fact that the particles created by the source (\ref{jdef}) have a
non-thermal spectrum weighted towards high energies:  it falls off roughly as $\omega_k^{-5}$ compared
with the exponential fall-off of the thermal distribution.  It is on account of this difference that
the integral equation (\ref{deltaneqn}) fails to respect the energy sum rule, and thus to be
self-consistent.  Comparing figures \ref{fig4} and \ref{fig6}, it appears that, although the created particles
are not thermalized instantaneously, a significant amount of thermalization has taken place
during the quarter-cycle between $t=10.2$ and $t=19.9$.  This appearance is somewhat illusory, however, as
we explain in section \ref{discuss}.
\subsection{Numerical details\label{numeric}}
The calculations reported above were performed using modest computing resources (a PC), and a moderate
numerical accuracy seems to be adequate for the issues we seek to address.  The most demanding aspect
of the calculations is the solution of the evolution equations (\ref{dndt}) and (\ref{dnudt}).  To
deal with these, we computed the evolution of $N$ modes (that is, $N$ values of the momentum $k$) for the
distribution functions $n_k$ and $\nu_k$, with an upper cutoff $k_\mathrm{max}=200$.  Of the various
integrals that are needed, the energy integral (\ref{energy}) attaches the largest weight to
high-momentum modes and the chosen cutoff is large enough to make this integral reasonably accurate,
as illustrated by the integrands in figures \ref{fig4} and \ref{fig6}.  Since the calculations are
quite lengthy (the data needed to produce figure \ref{fig2} were obtained from 3 weeks continuous running)
we did not attempt to make the entire process converge by increasing $k_\mathrm{max}$ and $N$
systematically.  However, the results presented here, using $N=400$, which adequately resolves the
structure of the distributions generated, differ negligibly from those obtained using $N=200$, and
seem to be insensitive to the cutoff $k_\mathrm{max}$.

\begin{figure}
\begin{center}
\includegraphics[bb=0 0 212 283,scale=1.4]{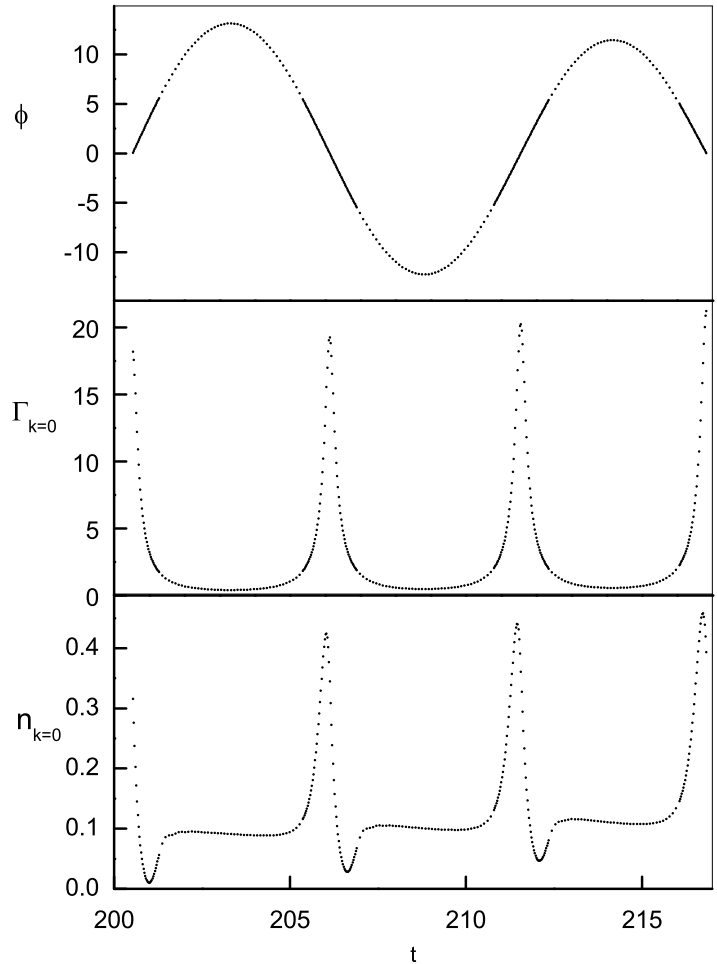}
\end{center}
\caption{\label{fig7}Variation of $\phi$, $n_{k=0}$ and $\Gamma_{k=0}$ over a typical cycle
of oscillation.  Data points are separated by the time interval $\Delta t$ explained in the text.}
\end{figure}

It is apparent from (\ref{dnudt}) that the functions $\nu_k(t)$ oscillate with frequencies roughly
equal to $\omega_k$.  The basic time stepper used was a semi-implicit method, which is stable for
these oscillations, with a fixed time step $\delta t$ about one-tenth of the period of the fastest
oscillation.  However, computation of the scattering integrals (\ref{alphadef}) and (\ref{gammadef})
is very time-consuming (even after analytical reduction to triple integrals), so $\alpha_k$ and
$\Gamma_k$, which change quite slowly relative to the fast oscillations, were updated at longer
time intervals $\Delta t$, using a predictor-corrector method that typically converged after two or
three iterations.  This larger time interval $\Delta t$ was assigned values between $20\delta t$
and $150\delta t$, according to the rate of change of $\Gamma_k$ (see figure \ref{fig7}). The values
of $n_k$ needed in these integrals were obtained by interpolating the $N$ tabulated values with a
cubic spline.

The quality of the data obtained by these methods is illustrated in figure \ref{fig7}, which shows
the variation over a typical cycle of $\phi$ and the $k=0$ modes of $\Gamma_k$ and $n_k$.  The
sharp peaks in $\Gamma_k$ appear in all modes, and are largely due to the factor $1/\omega_k$ in
(\ref{gammapresc}).  The basic operations of evaluating integrals, and so on, were carried out with
a tolerance of $0.5\%$ or better.  The energy (\ref{energy}), which formally is exactly conserved,
exhibits small variations, but is constant to within about $1\%$, which we estimate as the
overall accuracy of the calculation.
\section{Summary and discussion\label{discuss}}
According to quantum field theory, the equation of motion for the expectation value of a scalar
field is a complicated, non-local integro-differential equation.  In applications to inflationary
cosmology, it is typically assumed that this equation of motion can be replaced by a
local equation similar to (\ref{localeom}).  If it is further assumed that the state of the
quantum system is fairly close to thermal equilibrium, then it seems plausible that one should
be able to apply the methods of linear response theory, which yield quantitative estimates of
the friction coefficient $\eta(\phi)$ and the effective potential $V_\mathrm{eff}(\phi)$.

Here and in \cite{lawrie2002,lawrie03} we have tried to address the question whether the local equation
of motion obtained in this way adequately approximates the actual non-equilibrium evolution, at least
when this evolution is sufficiently slow.  The indications are that it does not, but the treatment we
have given cannot be considered as definitive, and we wish to set out clearly the issues that are involved.

The first issue is that we have treated the non-equilibrium evolution on the basis of an \textit{approximate}
set of local equations (\ref{phieom}), (\ref{chisq}), (\ref{dndt}) and (\ref{dnudt}).  As described in detail
in \cite{lawrie03}, they arise from a local \textit{Ansatz} for quasiparticle self-energies, and it is
not easy to assess how good an approximation this is.  For the purposes of estimating the non-equilibrium
behaviour numerically, other approximation schemes have recently become available (see, for example,
\cite{aarts,berges03,berges04,bergesserreau04}) which are in some respects superior to this one. However
for the purpose of deriving a local effective equation of motion, approximations more or less
equivalent to ours would seem to be inevitable.  To be precise, the question we address here is whether
the fully local equation of motion (\ref{eomadiab}) as derived from linear response theory yields a
good approximation to the non-equilibrium behaviour as described by our evolution equations.  If it
does not (as we seem to find), then it is also unlikely to provide a good approximation to the true
non-equilibrium dynamics.

The phrase ``fully local'' above means the following.  The evolution equations (\ref{phieom}),
(\ref{dndt}) and (\ref{dnudt}) are local, in the sense that they involve $\phi(t)$, $n_k(t)$ and $\nu_k(t)$
together with their time derivatives, all evaluated at the same time $t$:  they describe a Markovian
process.  In principle, one can envisage solving (\ref{dndt}) and (\ref{dnudt}) for $n_k(t)$ and
$\nu_k(t)$ as functionals of $\phi(t)$.  These solutions are non-local: they depend on the entire
history of $\phi(t)$. So, substituting these solutions into the remaining equation (\ref{phieom}), we
have a non-local equation of motion for $\phi$.  This non-local equation might be approximated by the
``fully local'' equation (\ref{eomadiab}) if the quantity $g\phi\langle\chi^2\rangle$ has a time-derivative
expansion whose first two terms are $\Delta V'(\phi)$ and $\eta(\phi)\dot\phi$.  Although we cannot
find the complete functionals $n_k(t)$, and $\nu_k(t)$, we can attempt to find their time-derivative
expansions.  As explained in section \ref{eveqns}, these expansions do not exist.  However, if we make
the further approximation of calculating self-energies from an equilibrium particle distribution rather
than the non-equilibrium distribution, we \textit{can} obtain a time derivative expansion, which essentially
reproduces the results of linear response theory.  This is what we have referred to as the adiabatic
approximation.  In section \ref{results}, we compared numerically the non-equilibrium evolution with
that resulting from the adiabatic approximation, and found very large differences.  We also saw that,
for the model studied here, thermalization is not very fast compared with the motion of $\phi$.  Intuitively,
it seems that this poor thermalization might account for the failure of the adiabatic approximation.

The second issue we want to discuss is, therefore, that of thermalization.  As indicated in section
\ref{od}, the relaxation time $\tau_k=1/\Gamma_k$ is roughly equal to $\phi/\dot\phi$ in the initial
state from which figure \ref{fig2} was generated.  On the basis of the adiabatic approximation, which
indicates strong overdamping, one would perhaps expect the non-equilibrium evolution to exhibit a period
of overdamped motion, albeit with significant corrections due to a modest thermalization rate.  What
we actually found was oscillatory motion with a period much shorter than $\tau_k$.  Over one quarter-period
of this motion, the particle distribution changes from that shown in figure \ref{fig4} to that shown in
figure \ref{fig6}.  From these figures, it would seem that significant thermalization has taken place
over a period of time much shorter than $\tau_k$.  However, figure \ref{fig7} suggests a different
explanation.  From that figure, we see that particles with a non-thermal distribution are radiated
over a short period of time as $\phi^2$ decreases towards 0, and that these particles, rather than being
thermalized, are quickly reabsorbed as $\phi^2$ subsequently increases.  So thermalization is indeed
slow on the time scale of the non-equilibrium motion.

We cannot altogether rule out the possibility that it is this slow thermalization that accounts for
the failure of the adiabatic approximation in the situation we have studied.  However, we do not think
that this is the correct explanation.  One reason is that, as we have just remarked, consideration of
the adiabatic approximation alone would suggest that thermalization, while modest, is not so slow as
to produce the gross discrepancy apparent in figure \ref{fig2}.  A more compelling reason is that, as
discussed in section \ref{od}, the non-equilibrium motion is not well represented by the adiabatic
equation (\ref{eomadiab}) even at those times when the state is close to thermal equilibrium. This
strongly indicates that the step of replacing, say, $\Gamma_k[n]$ with $\Gamma_k[n^{\mathrm{eq}}]$
is wrong in principle and not just because of the quantitative discrepancy which, as seen in figure
\ref{fig5}, may be quite small.
In more detail, suppose that the distributions $n_k(t)$ and $\nu_k(t)$ are indeed close to thermal
equilibrium.  Then  it should be a good approximation to linearize the evolution equations (\ref{dndt})
and (\ref{dnudt}), using $n_k(t)=n_k^\mathrm{eq}(t)+\delta n_k(t)$ and
$\nu_k(t)=\nu_k^\mathrm{eq}(t)+\delta \nu_k(t)$.  In particular, the linearized version of (\ref{dndt})
is
\begin{equation}
\partial_t\delta n_k+\int_0^\infty dk'K(k,k')\delta n_{k'}=\partial_tn^{\mathrm{eq}}_k
-\frac{\dot{\omega}_k}{\omega_k}
{\mathrm{Re}}\,[\nu_k^{\mathrm{eq}}+\delta\nu_k].
\label{linearbe}
\end{equation}
Unlike (\ref{deltaneqn}), this equation is self-consistent; the energy sum rule merely gives an expression
for the rate of change of the quantity $\int dk\,k^2\omega_k\delta n_k$. (This equation is not self-contained,
however.  To solve it, we would need a prescription, such as the conservation of energy, for determining
the time-dependent temperature that specifies $n_k^{\mathrm{eq}}$.  As a matter of fact, the scattering
integral $S_k$ vanishes when $n_k^\mathrm{eq}$ has an arbitrary chemical potential.  In the adiabatic
calculations of section \ref{results}, we did not allow for the possibility of a non-zero chemical potential
because there is no conserved particle number that could be used to determine its value.) The important
point here is that the linearized equation (\ref{linearbe}) is a \textit{differential equation}.  Its solution is
again a non-local functional of $\phi$, which will not reproduce the adiabatic equation (\ref{eomadiab}),
regardless of how small $\delta n_k$ might be.  To do that, we need not just a linear approximation, but
a time-derivative expansion which, as we have emphasized, is not self-consistent.

In summary, our numerical study indicates that that the non-equilibrium evolution of $\phi$ is not
well described by a local equation of motion with the friction term estimated by linear response theory.
It is possible that an adiabatic approximation would work better for a model in which thermalization is
more efficient, and we plan to explore this in future work.  However, for the reasons just discussed, we
do not think that poor thermalization is the principal cause of the large discrepancies we observed.


\end{document}